# Effect of very slow O diffusion at high temperature on very fast H diffusion in the hydride ion conductor LaH$_{2.75}$O$_{0.125}$

Yoyo Hinuma

Department of Energy and Environment, National Institute of Advanced Industrial Science and Technology (AIST), 1-8-31, Midorigaoka, Ikeda, Osaka 563-8577, Japan
*y.hinuma@aist.go.jp

**ABSTRACT:**
Neural network potential based molecular dynamics (MD) simulations on the excellent H conductor LaH$_{2.75}$O$_{0.125}$ show that O starts diffusing above a critical temperature of $T_c$~550 K, according to the variance of atom positions regardless of the time step. The original diffusion process at temperatures below $T_c$ has an activation barrier of 0.25 eV. Use of MD simulations with various O and La mass revealed, at above $T_c$, the coexistence of the 0.25 eV process and an additional diffusion process with an activation barrier of 0.20 eV. The O and La have strongly anharmonic characters.

**Keywords:**
Arrhenius plot, two activation barriers, LaH$_{2.75}$O$_{0.125}$, neural network potential, molecular dynamics simulations

## 1. Introduction

Ionic conductors can have exceptionally low activation barriers. Hydride ion conductors are useful for fuel cells, hydrogen storage, and hydrogen-based redox reaction control. BaTiO$_{3-x}$H$_x$[1], SrVO$_2$H (two-dimensional diffusion)[2], and LaSrCoO$_3$H$_7$ (mobile only above 675 K)[3] have very low H$^-$ activation barriers of ~0.3 eV or less. LaH$_{3-2x}$O$_x$ has a low activation barrier of 0.3-0.4 eV at $x<0.25$, the experimentally observed H$^-$ conductivity in LaH$_{2.8}$O$_{0.1}$ is extremely high at ~10$^{-3}$ S/cm near room temperature and ~0.03 S/cm at 400 K, and computational simulations suggest three-dimensional (3D) diffusion of H$^-$. Its conductivity is purely ionic; the transport number of (ionic conduction)/(ionic + electronic conduction) is reported as 99.98%. [4]

The Arrhenius plot of the experimental ionic conductivity of LaH$_{2.8}$O$_{0.1}$ appears to bend at ~350 K (Fig.2 of ref. [4]). A transition between two quasi-linear Arrhenius regimes of the self-diffusion coefficient at a critical temperature around ~440 K, instead of a single linear Arrhenius regime, cannot be ruled out based on a close inspection of neural network potential (NNP) based molecular dynamics (MD) simulation results of $LaH_{2.75}O_{0.125}$ (Fig.3(c) of ref. [4]), although this transition is not discussed at all in the original reference. [4] Arrhenius plot bending at a critical temperature was found in NNP-MD simulations of $Li_{0.33}La_{0.56}TiO_3$, $Li_3YC_6$, and $Li_7P_3S_{11}$,[5] and experimentally in $Na_{0.5}Bi_{0.49}TiO_{2.985}$[6], $BaZr_{0.5}In_{0.5}O_{2.25}H_{0.5}$[7], and $LaBi_{1.9}Te_{0.1}O_{4.05}Cl$[8].

This paper used NNP-MD simulations to clarify the physics behind the transition between quasi-linear Arrhenius regimes in $LaH_{2.75}O_{0.125}$ (hereafter LaHO). LaHO is experimentally known to have a high H conductivity, bending of its Arrhenius plot that is experimentally found and computationally possible though not confirmed, and its crystal structure is very simple with just three atom sites, namely a cation site, a tetrahedral anion site, and an octahedral anion site. Deriving a sophisticated potential model of LaHO, which was already discussed in detail in Refs. [4,9,10], and making efforts toward better prediction of absolute values of diffusion coefficients and $T_c$ are outside the scope of this paper.

Atom mass change was used here as a convenient artificial procedure to change the speed of atom movement without touching the potential energy landscape. Hydrogen mass repartitioning[11] is a trick used in MD simulations of organic molecules where details of C-H bond vibrations are not important yet their high frequencies forces use of a small time step. Moving mass from heavier atoms in the framework, such as C, to H decreases these frequencies, allowing use of larger time steps. Mass was not repartitioned in this study but the mass of all elements, or O and La only, were changed.

Figure 1 illustrates how atom mass $m$ and temperature $T$ affects the diffusion of an atom. The Born-Oppenheimer approximation makes the potential landscape $U$ independent of $m$. The mean, over atoms, of the variance of the displacement of each atom with regard to the equilibrium point, $\langle \Delta r^2 \rangle = \langle Var_n(\Delta r_{in}) \rangle_i = \langle Var_n(\Delta x_{in}) + Var_n(\Delta y_{in}) + Var_n(\Delta z_{in}) \rangle_i$ (see left half of Figure 1), is determined by $U$ and $T$. Here, $\Delta x_{in}$, $\Delta y_{in}$, and $\Delta z_{in}$ are the displacements in the $x$, $y$, $z$ directions, respectively, of atom $i$ at step $n$, and $\Delta r_{in}^2 = \Delta x_{in}^2 + \Delta y_{in}^2 + \Delta z_{in}^2$. The variances are calculated over all $n$ for each atom $i$, which is stressed with the subscript $n$ after $Var$. The equilibrium point for each atom does not need to be obtained explicitly. In fact, the reference point of displacements can be anything because the variance of displacements of a given atom is equal to the variance of the atom position. This contrasts the dependence on the initial positions when deriving the tracer diffusion coefficient, where the displacement is taken against an initial position. In non-diffusing atoms, atoms simply oscillate around the equilibrium position and $|\Delta r_{in}|$ cannot be larger than few 0.1 Å regardless of $n$, and $<\Delta r^2>$ quickly converges to a constant value on the order of 0.01 Å$^2$ as the simulation time increases. In contrast, $|\Delta r_{in}|$ of diffusing atoms can exceed few 0.1 Å after a sufficiently long time as no upper limit exists on $|\Delta r_{in}|$, and $<\Delta r^2>$ increases with simulation time. The simple but often overlooked quantity $<\Delta r^2>$ serves as a robust indicator on whether an atom can diffuse or not, even for very slowly diffusing atoms where the diffusion coefficient cannot be reasonably obtained.

## 2. Computational procedure

MD simulations were conducted using the commercially available Matlantis package from Preferred Networks with the universal PreFerred Potential (PFP) version 4.0.0 [12], a NNP trained on the Perdew-Burke-Ernzerhof (PBE) generalized gradient approximation (GGA) to density functional theory[13]. LaHO has a fluorite-based structure and one out of 24 anion sites is intrinsically vacant. The canonical, or constant number of atoms, volume, and temperature (NVT), ensemble, and a Nosé–Hoover thermostat[14,15] were used. The time step was 1 fs, as in Ref. [4], and 1,000,000 steps (1 ns) were simulated. The employed MD supercell (Fig. 2) was a 2×2×3 supercell of the NNP MD-based primitive cell in Table S3 of Ref. [4] (lattice parameters $a=b=3.987$ Å and $c=5.670$ Å for La$_{16}$H$_{44}$O$_2$). The basis vectors were taken parallel to the $x$, $y$, and $z$ directions. The H/O/vacancy ordering in Fig. S4 of Ref. [4] was adopted as the starting point of MD calculations. The simulation duration was shortened and the supercell was downsized from Ref. [4] to reduce computational costs. The universal PFP is easy to access because the user does not need to, and cannot, train the potential, but is slower than a dedicated, fine-tuned potential.

The tracer diffusion coefficient in the $x$ direction, $D_x$, was obtained by plotting the mean square displacement (MSD) in the $x$ direction, $\langle \Delta x_{in}^2 \rangle_i$, versus $n\tau$ and fitting to $\langle \Delta x_{in}^2 \rangle_i = 6D_x n\tau$. Here, $\Delta x_{in}$ is the displacement of atom $i$ in the $x$-direction at step $n$ from the initial position at $n=0$, $<\cdot>_i$ indicates taking the mean of all atoms $i$, and $\tau$ is the time step. The value of $\langle \Delta x_{in}^2 \rangle_i$ depends on $n$. Diffusion coefficients along the $y$ and $z$ directions, $D_y$ and $D_z$, respectively, were obtained similarly, and the 3D diffusion coefficient is $D=D_x+D_y+D_z$.

## 3. Results and discussion

Figure 3 shows the MD results when all atom masses were changed to $M$=1, 2, 4, and 16 times the natural values (×1 to ×16, respectively). Simulations were conducted at 0.0001 K$^{-1}$ intervals of $1/T$ (0.00005 K$^{-1}$ for ×1). Fig. 3(a) gives Arrhenius plots of log $D$ versus $1/T$. The ×1 plot has two quasi-Arrhenius regimes transitioning at a critical temperature of $T_c \approx$ 550 K ($1/T_c \approx$ 0.00182 K$^{-1}$). The activation barriers from the linear regressions for $T<T_c$ and $T>T_c$ are 0.25 and 0.20 eV, respectively, and the former is in line with the experimentally observed 0.3-0.4 eV[4]. The lower $D$ of the two regressions is denoted as $D_L$ (L for "lower"), and Fig. 3(b) plots log $D$ – log $D_L$ versus $1/T$. The $D$ with atom mass increased $M$ times is roughly -0.5 log $M$=-0.15 log$_2$ $M$ below log $D_L$.

The scaling relations between mass $m$, atom speed $v$, and $D$ are discussed here. Variables with subscript 0 are considered reference values. Fig. 3(b) indicates that $(m/m_0)^{-1/2} \approx D/D_0$ holds over all temperatures and that $T_c$ is the same regardless of $M$. From another viewpoint, scaling all masses proportionately simply changes the time scale of the simulation. When $m$ is increased from $m_0$ to $m$, $v/v_0 \approx (m/m_0)^{-1/2}$ because $mv^2/2 \approx 3k_BT/2$ and $v \approx \sqrt{3k_BT/m}$ ($k_B$ is the Boltzmann constant). Doubling the speed of an atom doubles the opportunity for a random walk, or in other words, there are two random walk chances instead of one over a certain time period. As $\langle \Delta x_{in}^2 \rangle_i = 6D_x n\tau$, doubling $v$ means that reaching $\langle \Delta x_{in}^2 \rangle_i$ with the same number of time steps $n$ is attained by halving $\tau$, thereby doubling $D_x$. Similar logic applies to the $y$ and $z$ directions, thus $D/D_0 \approx v/v_0$ hence $D/D_0 \approx (m/m_0)^{-1/2}$.

A larger mean square displacement can be attained over the same MD simulation time by decreasing $m$, but there is a catch. In principle, the oscillation period of a classic harmonic oscillator described by $m\ddot{x} = -kx$ should be sampled using the same number of MD steps $N$. The period of this oscillator is $2\pi\sqrt{m/k} = N\Delta t$, where $\Delta t$ is the appropriate time step to be used as $\tau$ in MD calculations. As $\tau/\tau_0 = \Delta t/\Delta t_0 = (m/m_0)^{1/2}$, which is also the same as $(D_x/D_{x0})^{-1}$, the quantity $6D_x\tau = \langle \Delta x_{in}^2 \rangle_i / n$ should be the same regardless of $m$ when the appropriate time step is used. In other words, the same diffusion distance should be covered using the same number of steps, thus scaling all masses cannot improve the quality of the derived $D$. This argument also holds in $y$ and $z$ directions and therefore in 3D.

Fig. 3(c) plots $\langle \Delta r^2 \rangle$ versus $T$ for La. La is always trapped near the equilibrium point. Points for different $M$ overlap for each $T$, which is a natural result as the potential is independent of $M$. Fig. 3(d,e) gives $\langle \Delta r^2 \rangle$ versus $T$ for O. When $M=1$, O is trapped near the equilibrium point below $T_c$ but starts diffusing above $T_c$. The actual transition temperature increases when $M$ increases, which may be an artifact of too short simulations. Interestingly, the $\langle \Delta r^2 \rangle$ versus $T$ plots for La and O (below $T_c$) have a linear relation but its intercept at $T=0$ is not $\langle \Delta r^2 \rangle = 0$. A simple harmonic oscillator with the standard Hamiltonian $\hat{H} = \left( -\frac{\hbar^2}{2m}\frac{d^2}{dx^2} + \frac{m\omega^2}{2}x^2 \right)$ is considered. Its eigenvalues are $E_n = (n+1/2)\hbar\omega$ and the corresponding variance of the position is $\langle x^2 \rangle = E_n/(m\omega^2)$. As $E \propto T$, $\langle x^2 \rangle \propto T$ is expected in a harmonic potential, and in 3D, $\langle x^2 \rangle + \langle y^2 \rangle + \langle z^2 \rangle \propto T$. This clearly does not hold in Figs. 3(c,d), thus the potentials for both O and La are strongly anharmonic.

There is rich literature on phonon-ion interactions, or the relations between lattice dynamics and ionic conductivity. Topics include relating ion polarizability to superionic conduction, softer lattices to low activation energy, specific phonon frequencies to activation barriers, average phonon frequencies and phonon band centers, and ion diffusion to rotational motion of polyanionic units (paddle wheel mechanism).[16] An empirical Meyer-Neldel rule exists stating that the logarithm of the prefactor is a linear function of the activation barrier where the prefactor decreases when the activation barrier is decreasing, which means that reducing the activation barrier does not necessarily increase the ionic conductivity [17-19].

Figure 4 plots the MSD in the $x$, $y$, and $z$ directions, which are $\langle \Delta x_{in}^2 \rangle_i$, $\langle \Delta y_{in}^2 \rangle_i$, and $\left\langle \Delta z_{in}^2 \right\rangle_i$, respectively, versus simulation time $n\tau$ for H, La, and O when $1/T=0.014$ K ($T=714$ K) and $M=1$. Results for H is given in Fig. 4(a). This is a prototypical result for a good ion conductor, where the MSD versus simulation time plot lies on a straight line. The slope is simply $6D_x$, $6D_y$, or $6D_z$, and the value of $D$ is the sum of $D_x$, $D_y$, and $D_z$. The average slope is roughly 160 Å²/ps, corresponding to $D=8\times10^{-6}=10^{-5.1}$ cm²/s. Fig. 4(b) shows results for La. The plot is flat, thus $\left\langle \Delta x_{in}^2 \right\rangle_i \approx \left\langle \Delta y_{in}^2 \right\rangle_i \approx \left\langle \Delta z_{in}^2 \right\rangle_i \approx (1/3)\left\langle \Delta r^2 \right\rangle$ over the entire simulation duration. It is possible to force a linear fit passing through the origin and obtain $D$, which is $2\times10^{-9}$ cm²/s. However, this $D$ has very little meaning other than "very small", especially as this value is strongly dependent on how much simulation time is considered and $D$ becomes zero when the simulation is infinitely long. In contrast, $<\Delta r^2>$ is a physically meaningful quantity that does not depend on the simulation time. Fig. 4(c) shows results for O. Hopping of atoms, which discretely changes $\left\langle \Delta x_{in}^2 \right\rangle_i$, $\left\langle \Delta y_{in}^2 \right\rangle_i$, and $\left\langle \Delta z_{in}^2 \right\rangle_i$, are sporadic, and the MSD versus simulation time plot cannot be reasonably described as a straight line. Here, O can diffuse, but the diffusion is very slow and $D$ cannot be obtained with high precision. A linear fit passing through the origin gives $D=3\times10^{-8}$ cm²/s, although this value serves as a rough guide only. The $D$ of O is two orders of magnitude smaller than H, thus LaH$_{2.75}$O$_{0.125}$ is an excellent H conductor and a poor O conductor, as expected.

The lower bound of a reliable $D$ is discussed below. A typical value of the nearest neighbor cation-anion bond distance in ionic compounds is ~2 Å, hence an average displacement of 1 Å, or MSD of 1 Å², over a MD simulation may be regarded as a reasonable arbitrary value required to obtain $D$ with acceptable precision. A practical NNP MD simulation can handle 1,000,000 steps with close to 1000 atoms in a MD cell. When the time step is 1 fs, the minimum reasonable $D$ is $(1/6)\times10^{-16}/10^{-9}=10^{-7.8}$ cm²/s. The minimum $D$ considered reliable in this study is $D = 10^{-6.5}$ cm²/s, which equates to a displacement of ~4.4 Å after a 1 ns simulation. A 100,000-step calculation of about 100 atoms in the MD cell is near the practical limit of first-principles MD simulations, and the number of atom hopping events over the simulation duration is roughly two orders of magnitude smaller than a NNP-MD simulation. The minimum reasonable $D$ from first-principles MD assuming an MSD of 1 Å² over the simulation becomes $(1/6)\times10^{-16}/10^{-10}=10^{-6.8}$ cm²/s, one order of magnitude higher than NNP-MD. One may be forced to compromise with a smaller MSD, though the reliability of $D$ becomes even worse. Evaluations of $D$ using such NNPs were very limited owing to the difficulty of obtaining good NNPs. The universal PFP was made available only as recently as 2021.

Scaling masses of all atoms by the same amount simply resulted in a constant shift of log $D$, but what happens when the mass of H, which diffuses over the entire studied temperature range, is not changed but the masses of La and O are? Figure 5 shows the MD results when masses of La and O were changed to $M$=0.25, 1, 4, 16, 64, 256, and 1024 times the natural values (×0.25 to ×1024, respectively). The values of $D$ for ×1 and its two regression lines are the same as in Fig. 3.

Arrhenius plots are shown in Fig. 5(a). The H mass was not changed, so a naïve guess is that H can diffuse at the same speed regardless of $M$ and thus $D$ is almost the same for a given $T$, but this is totally wrong. The values of log $D$ minus the linear regression of $M$=1 at $T<T_c$, denoted as log $D_{R\_lowT}$ (regression-low temperature), is shown in Fig. 5(b). The $<\Delta r^2>$ versus $T$ plots for La and O in Figs. 5(c-e) show the same trends as in Fig. 3(c-e).

Fig. 5(b) is discussed in detail below. When $T<T_c$, the points lie on horizontal lines, and the lines for points with $M$ other than 0.25 are almost evenly spaced. The horizontal dashed lines are drawn at 0.075 intervals as a guide. The diffusion coefficient in this region can be written in the form $D_{lowT}(M,T) = D'_{lowT}(M)\exp\left(-\frac{E_a}{k_B T}\right)$, or $\log D_{lowT}(M,T) = \log D'_{lowT}(M) - \frac{E_a}{k_B T}$, because the slope of $\log D_{lowT}$ has no $M$ dependency (Fig. 5(b)) and $E_a$ does not depend on $M$. Although there is no clear justification, $\log D'_{lowT}$ can be further decomposed into $\log D'_{lowT}(M) = \log D''_{lowT} - \alpha \log M$, where $D''_{lowT}$ and α are constants independent of $M$ and $T$.

$E_a$ is the energy where the non-diffusing atoms are at optimum positions for letting the diffusing atom pass through. This $E_a$ does not depend on $M$, as the optimum atom positions are based on the potential only and not on $M$. Path-blocking non-diffusing atoms must make way in a short timeframe, in the order of the reciprocal of the attempt frequency, otherwise the diffusing atom cannot hop along the path. Cooperation of atoms at the bottleneck to allow diffusion may be regarded as implicit concerted diffusion. The velocity of non-diffusing atoms is affected by $M$, and heavier O are less likely to complete necessary movement that enables diffusion in the allowed timeframe, which results in a lower prefactor when O is heavier.

When $T>T_c$, points for different $M$ line up on slanting lines parallel to each other, thus the activation barrier $E_b$ in this regime is the same regardless of $M$. O starts to diffuse very slowly, or at least has energy above a certain threshold that allows sufficiently frequent displacement from the equilibrium position, at $T>T_c$. I speculate that this lower energy barrier arises from a missing O at the bottleneck. Missing O can only happen when the binding of O from the equilibrium position becomes weaker than a certain threshold corresponding to the critical temperature. There are only two activation barriers over the whole range of $T$ and $M$, suggesting that the $E_b$ is caused by a discretely changing mechanism, such as the number of O at the bottleneck, rather than a continuously changing mechanism where the activation barrier can take a range of values.

To understand how $D$ changes at $T>T_c$, a hypothetical system is considered with two diffusion paths between two sites, A and B, with prefactors $D_1$ and $D_2$ and activation barriers $E_1$ and $E_2$, respectively (Fig. 6). The net diffusion coefficient can be given as $D = D_1 \exp(-E_1/k_B T) + D_2 \exp(-E_2/k_B T)$. If $D_1 \gg D_2$ and $E_1 > E_2$, this is a downward convex curve where the first and second term dominates at small and large $1/T$, respectively, resulting in effective activation barriers of $E_1$ and $E_2$, respectively. There is a crossing temperature where the two terms are the same.

In the diffusion of LaHO, the crossing temperature $T_{c2}$ decreases, or $1/T_{c2}$ increases, as $M$ increases, and $T_c \approx T_{c2}$ at $M=1024$. Controlling $T_{c2}$ was attained through simulations changing the masses of (almost) immobile species. The $\log D - \log D_{R\_lowT}$ takes the same value independent of $T$ above $T_{c2}$, hence the activation barrier above $T_{c2}$ is effectively $E_a$.

The diffusion phenomenon of LaHO may be described as follows. At $T<T_c$, there is a single diffusion process in the form $D = D_0 \exp(-E_a/k_B T)$, while at $T<T_c$, there is a coexistence of two diffusion mechanisms resulting in $D = D_1 \exp(-E_a/k_B T) + D_2 \exp(-E_b/k_B T)$. The relations $D_2<D_1$ and $E_b<E_a$ are in line with the Meyer-Neldel rule. The quantity $D_1/D_0$, which is the ratio of activation barrier $E_a$ that is not blocked at $T>T_c$, represents the effect of enhanced O movement above at $T>T_c$. The very small number of O atoms compared to H means that not all $E_a$ paths are affected by O. This $D_1/D_0$ increases toward unity ($\ln D_0 - \ln D_1$ approaches zero) as O and La becomes heavier ($M$ becomes larger) and O movement decreases. The coexistence of two processes could not be identified, thereby resulting in an inaccurate description, if calculations with large $M$ were not conducted.

Phonon modes at very low energies are found in Na phonon DOS of Na conductors $Na_3ZnGaS_4$ and $Na_3ZnGaSe_4$ [20] and in all Li, O, and Si phonon DOS in amorphous $Li_2Si_2O_5$, a good Li conductor, but not in crystalline $Li_2Si_2O_5$, which is a bad conductor [21]. The amorphous structure has a higher Li conductivity than the crystalline structure in $LiAlO_2$. [22] Freezing of some atoms inhibit diffusion by killing anharmonic low energy phonon modes. For example, freezing Se stops diffusion of Ag in $Ag_8SnSe_6$ [23], freezing $PS_4$ decreases the Na diffusion constant by a factor of five in $Na_3PS_4$ [24], and freezing the host decreases the Cu diffusion constant by a factor of four in $Cu_7PSe_6$, respectively [25].

Calculation of phonon modes and DOS is an established but difficult procedure, which was avoided in this paper. In contrast, $<\Delta r^2>$, as a constant quantity independent of simulation time, is very easy to obtain though information on how specific phonon modes affect diffusion cannot be extracted. Where to freeze atoms is not a trivial problem when non-diffusing atoms are, for example, in a double-well potential situation. Increasing the masses can attain reduced movement while not worrying too much about their explicit positions because the mass-increased atoms can move to suitable positions. The strong coupling between H diffusion and lattice vibrations of the host lattice in LaHO was confirmed through $D$ values from MD simulations with various O and La mass, without the need for complicated phonon calculations.

A downward convex curve is problematic because extrapolation from MD calculations at high temperature overestimates $D$ below the crossing temperature, and extrapolation from low temperature using activation barriers obtained using nudged elastic band (NEB) calculations at 0 K also overestimates $D$ above the crossing temperature. Note that an upward convex curve cannot be described using in the form

$$D = D_1 \exp(-E_a/k_B T) + D_2 \exp(-E_b/k_B T).$$

## 4. Summary

The diffusion mechanism of LaHO was identified with NNP-MD calculations. Two often overlooked approaches were used, namely use of the variance of atom positions regardless of the time step, $<\Delta r^2>$, to judge whether an atom can diffuse or not, and changing the mass of some, but not all, elements. The former finds very slow diffusion of atoms where the diffusion constant cannot be reliably obtained. These approaches were necessary to confirm that a critical temperature $T_c$ exists; when $T<T_c$, O cannot diffuse and H diffuses with an activation barrier of 0.25 eV, and when $T>T_c$, O can diffuse and H diffuses using two different processes with activation barriers of 0.25 eV (same as when $T<T_c$ where O cannot diffuse) and 0.20 eV, respectively. The latter approach was necessary to confirm the coexistence of two processes. The strong anharmonicity of O and La and its coupling to H may contribute to the unique ionic conductivity characteristics of LaHO.

**Acknowledgements**

This paper is based on results obtained from a project, JPNP21016, commissioned by the New Energy and Industrial Technology Development Organization (NEDO). The VESTA code[26] was used to draw Fig. 1.

**Conflict of interest statement**

The authors have no conflicts to disclose.

**Data availability statement**

The data that support the findings of this study are available from the corresponding author upon reasonable request.

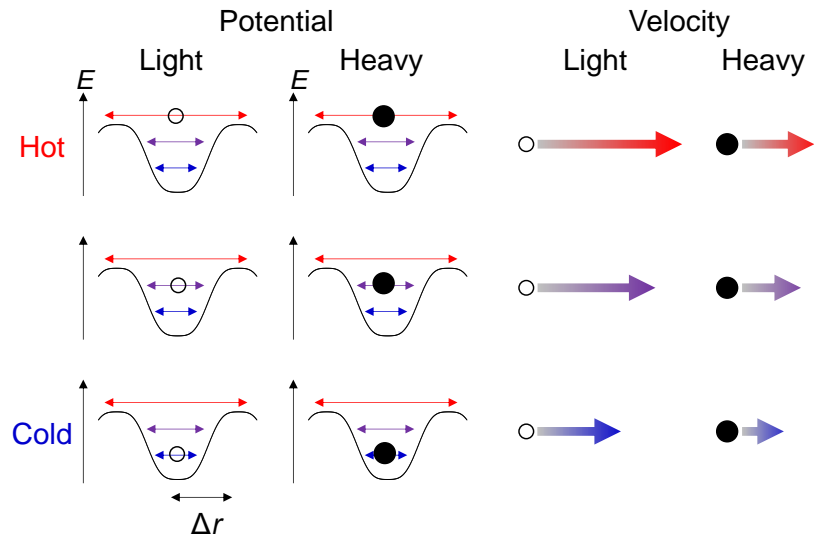

Fig. 1. Schematic on how the temperature (cold to hot) affects the position and velocity of atoms with different mass (light and heavy mass shown as white and black circles).

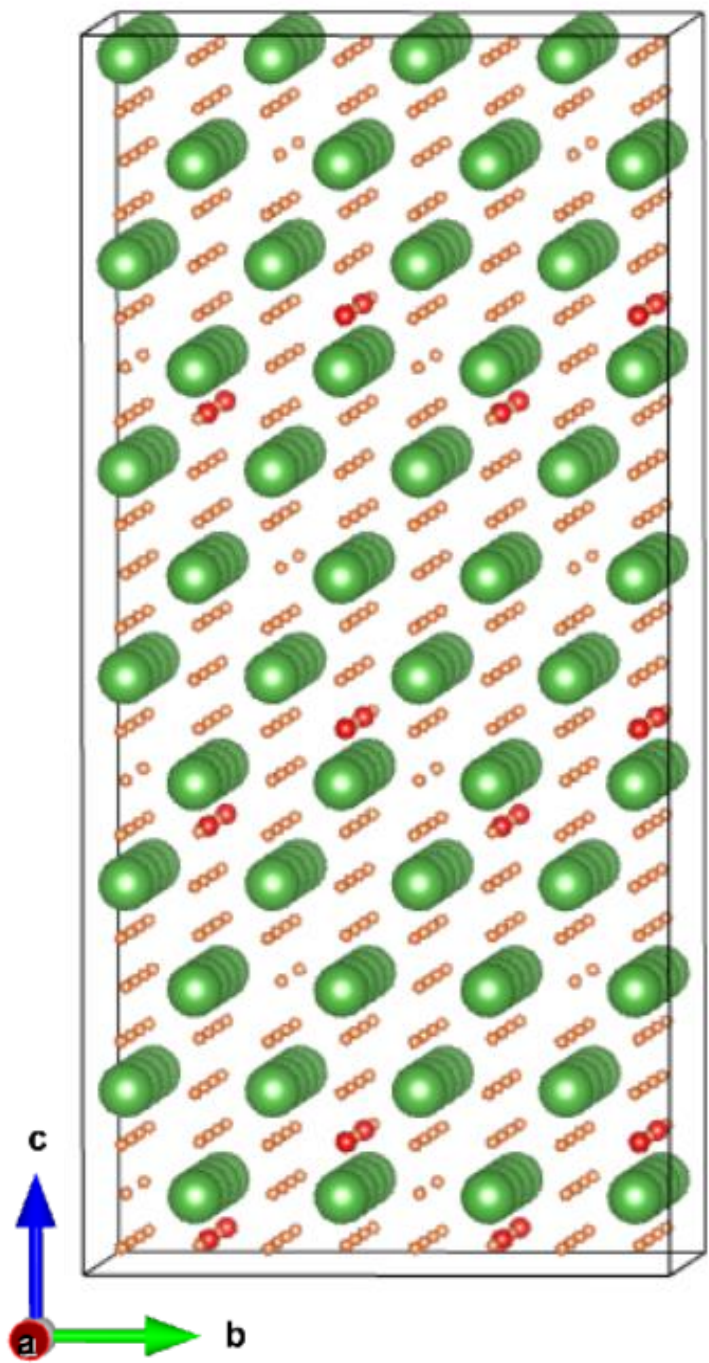

Fig. 2. LaHO supercell used in this study. La, O, and H are shown as very large green, large red, and small pink balls, respectively.

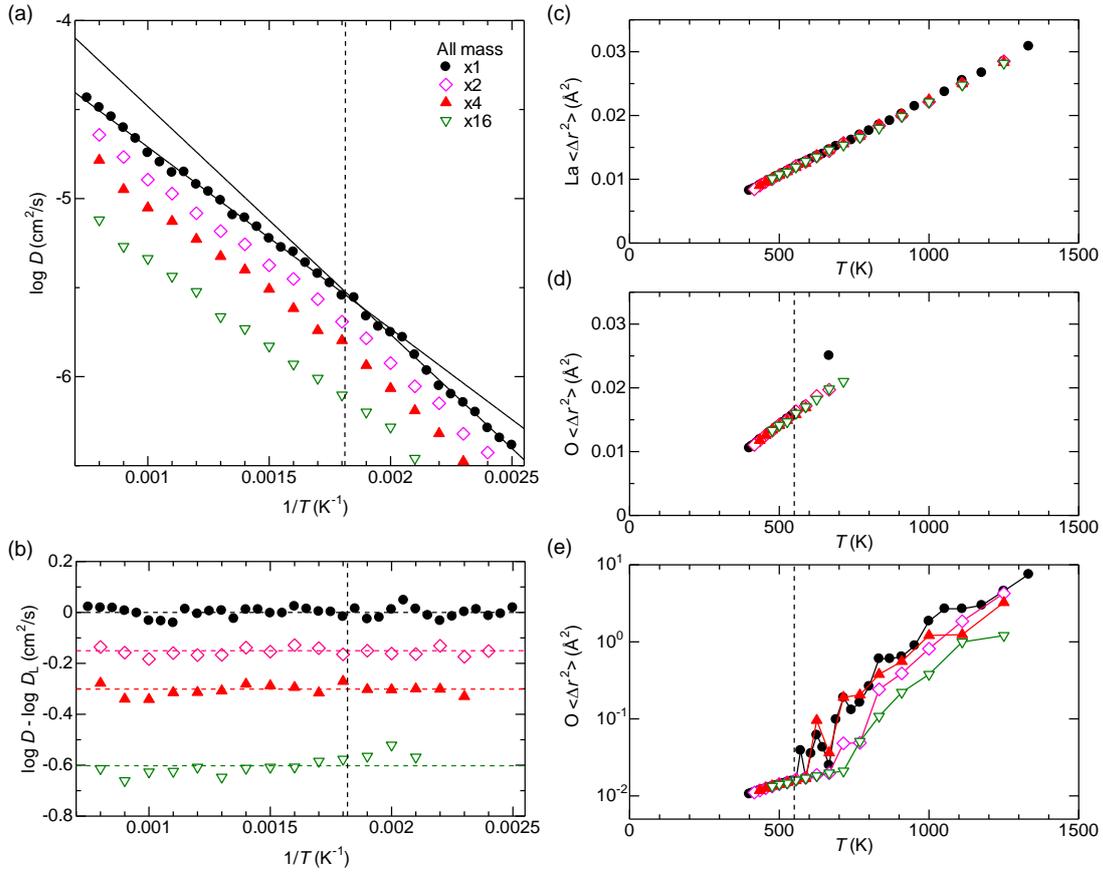

Fig. 3. (a) Arrhenius plot of log $D$ vs $1/T$ when all masses were multiplied by ×1 to ×16. The straight lines are regressions of ×1 results above or below $T_c$=550 K. (b) The vertical axis is log $D$ - log $D_L$, where log $D_L$ is the lower of the two regression lines in (a). Horizontal dashed lines are drawn at log(2)/2 or log(4)/2 intervals. The $<\Delta r^2>$ regardless of the time step of (c) La and (d) O, respectively. (e) The $<\Delta r^2>$ of O shown as a semilog plot. Lines are drawn to guide the eye. $T_c$=550 K is indicated as a vertical dashed line in (a,b,e).

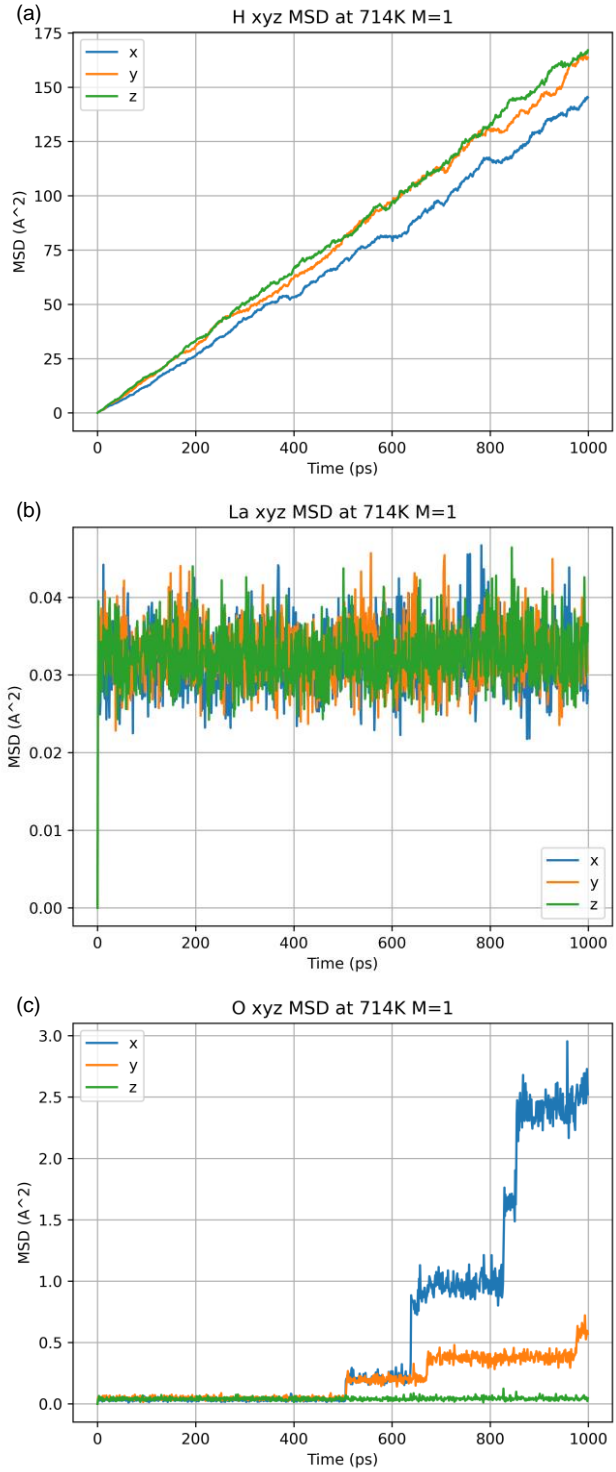

Fig. 4. Plots of the MSD in the three directions, $\langle \Delta x_{in}^2 \rangle_i$, $\langle \Delta y_{in}^2 \rangle_i$, and $\langle \Delta z_{in}^2 \rangle_i$, versus simulation time $n\tau$ for (a) H, (b) La, and (c) O when $1/T$=0.0014 K$^{-1}$ ($T$=714 K) and $M$=1.

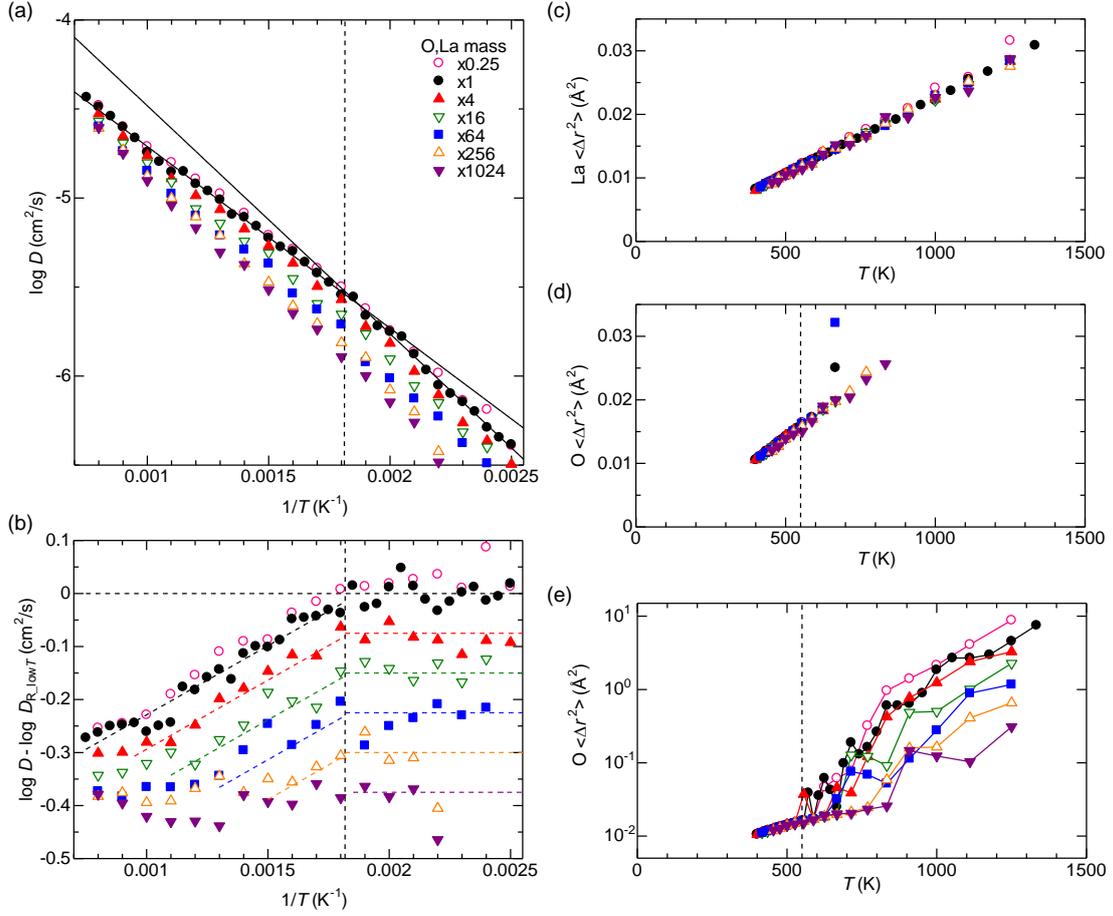

Fig. 5. (a) Arrhenius plot of log $D$ vs $1/T$ when O and La masses were multiplied by ×1 to ×16. The straight lines are regressions of ×1 results above or below $T_c$=550 K. (b) The vertical axis is log $D$ - log $D_{R\_lowT}$, where log $D_{R\_lowT}$ is the regression line for $T<T_c$ in (a). Slanted dashed lines are drawn parallel to each other, implying a common activation barrier. The $<\Delta r^2>$ regardless of the time step of (c) La and (d) O, respectively. (e) The $<\Delta r^2>$ of O shown as a semilog plot. Lines are drawn to guide the eye. $T_c$=550 K is indicated as a vertical dashed line in (a,b,d,e).

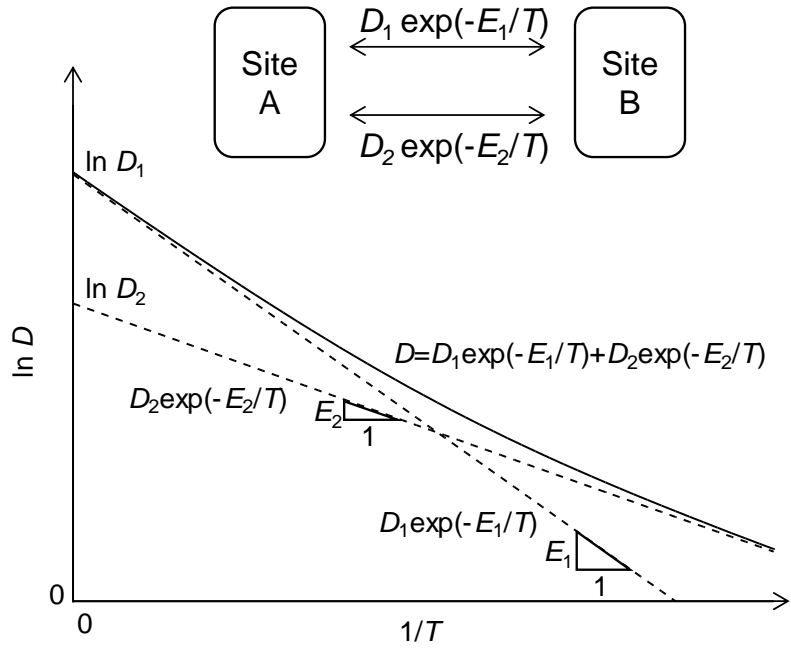

Fig. 6. Schematic of ln $D$ when the diffusion between sites A and B can happen using two independent channels with different prefactors and activation barriers. The value $T$ in this figure is the temperature times the Boltzmann constant. The two diffusion channels satisfy $D_2 < D_1$ and $E_2 < E_1$.